\pdfoutput=1
\documentclass[12pt]{article}
\usepackage[utf8]{inputenc}
\usepackage[T1]{fontenc}
\usepackage{hyperref}
\usepackage{longtable}
\usepackage{booktabs}
\usepackage{geometry}
\usepackage{array}
\usepackage{xcolor}
\usepackage{colortbl}
\usepackage{enumitem}
\usepackage{verbatim}
\usepackage{amsmath,amssymb}
\usepackage{tikz}
\usepackage[normalem]{ulem}
\usetikzlibrary{arrows.meta}
\newcolumntype{L}[1]{>{\raggedright\arraybackslash}p{#1}}
\newcolumntype{C}[1]{>{\centering\arraybackslash}p{#1}}
\definecolor{headerblue}{HTML}{1a1a2e}
\definecolor{rowgray}{HTML}{f9f9f9}
\newcommand{\code}[1]{\texttt{#1}}
\newcommand{\func}[1]{\texttt{#1}}

\title{GLP Networking API Specification}
\author{%
  \parbox[t]{0.45\textwidth}{\centering Dan Bachar\\[0.5ex]
    \small Technical University of Munich \\ and Technion--Israel Institute of Technology}%
  \hfill
  \parbox[t]{0.45\textwidth}{\centering Ehud Shapiro\\[0.5ex]
    \small London School of Economics \\ and Weizmann Institute of Science}%
}
\date{}
\begin{document}
\maketitle

\begin{abstract}
\noindent Multiagent deterministic GLP (madGLP)~\cite{shapiro2026implementing} runs Grassroots Logic Programs (GLP)~\cite{shapiro2025glp} as agents that communicate solely by asynchronous message passing, and assumes fair delivery: every message sent is eventually delivered.  Each agent runs on a person's smartphone and is named by the person's public key.  A grassroots platform depends on no global resource other than the network~\cite{shapiro2023grassrootsBA,shapiro2025atomic}, so the agents must reach one another face to face over Bluetooth Low Energy, on a local network, and across the Internet from behind NAT, with rendezvous servers that anyone can run for connection establishment.  What madGLP requires of such a network, and what the network requires of the platform above it, has not been specified.

Here, we specify the networking layer beneath madGLP. The guiding principle is that anything that can be done in GLP is done in GLP: the layer provides what GLP cannot reach by itself --- the network stack, the platform's location service, the platform's attestation of the application it runs, and cryptographic primitives --- and everything above that line is GLP code: the social graph, friendship, introduction, the distribution of addresses and the reconnection of peers whose address changed, and the rendezvous protocol.  

We specify three transport underlay realizations, all following one interface. Over the PAN and LAN transport underlays, an agent advertises a rotating recognition token derived from its key and the current time slot, so that a peer who knows the key recognises it before any contact and nobody else can track it; whether an agent answers an unrecognised contact is a trust level GLP sets per underlay, raised on entering a place GLP declared and lowered on leaving it.  Over the WAN underlay, a rendezvous server is a host with a static address running the headless layer and a GLP rendezvous agent, which observes addresses and coordinates hole punching, relays no payload, and belongs to no social graph.  On every underlay, the peers establish a mutually authenticated Noise session from their Ed25519 keys, IK when the initiator already knows the responder's key and XX only for a first contact under Open trust, and then attest to each other, per session and bound to the session's handshake hash, that each runs the application published under its developer's key, verified against Apple's and Google's public roots with no third party contacted during a session; we state what each platform's attestation proves and what it does not, and the runtime reports to the program whether a peer is attested.  One message protocol carries payloads on every underlay, over UDP datagrams and over GATT writes, with fair, unordered, atomic and authenticated delivery, which is what madGLP assumes.

\end{abstract}

\section{Introduction}
\label{sec:introduction}
Multiagent deterministic GLP (madGLP)~\cite{shapiro2026implementing} is a distributed runtime for Grassroots Logic Programs (GLP)~\cite{shapiro2025glp} in which autonomous agents communicate solely by message passing, each agent on a person's smartphone and named by the person's public key. This document specifies the networking layer beneath madGLP: the interface that enables madGLP agents to run on separate devices communicating peer-to-peer, and the layer functions behind the system predicates by which GLP programs reach it.  madGLP and the runtime's side of the seam are specified in the implementation paper~\cite{shapiro2026implementing}; the seam predicates are in the language paper's catalogue~\cite{shapiro2025glp}; the operating system above the runtime, the Grassroots Super-App with its social graph (in preparation), is what decides whom to befriend and what an attested or unattested peer may be used for.

\paragraph{Layering.}
The guiding principle is that anything that \emph{can} be done in GLP is done in GLP: the networking layer provides only what GLP cannot reach by itself---the network stack (PAN, LAN and WAN), the platform's location service (geofencing), the platform's remote attestation of the application it runs, and cryptographic primitives (Ed25519 signing and verification, the Noise handshake). Everything above this line is GLP agent code.  In particular, the social graph~\cite{shapiro2025atomic}, friendship, introduction, the distribution of addresses, the reconnection of peers whose address changed, and the rendezvous protocol are implemented entirely in GLP and are \emph{not} part of this specification; to the networking layer, all madGLP messages are opaque payloads.  The interface between the two consists of the API defined here together with ten system predicates of the language~\cite{shapiro2025glp}, whose layer functions Section~\ref{sec:system-predicates} specifies.

\paragraph{One overlay, three underlays.}
The API is underlay-agnostic at its core: madGLP issues the same calls regardless of how a message travels. Three underlays realize the core, the overlay being the one address space of public keys over them:
\begin{itemize}
\item \textbf{PAN} (Bluetooth Low Energy, Section~\ref{sec:ble}): communication between devices in direct radio range, enabling offline, infrastructure-free operation. The PAN is the bootstrap underlay where agents first meet face-to-face.
\item \textbf{LAN} (a local network, Section~\ref{sec:lan-discovery}): communication between devices on the same local network, found by DNS-SD over multicast DNS and carried over IP, with no radio and no server.
\item \textbf{WAN} (Internet Protocol, Section~\ref{sec:ip}): communication between remote devices at arbitrary distances, using UDP datagrams, with rendezvous servers for signaling and NAT traversal.
\end{itemize}
Some control operations are inherently underlay-specific---observing the agent's own public address is IP-only---and are specified with their underlay rather than in the agnostic core, and called on that underlay's sub-object (Section~\ref{sec:instantiation}). PAN and LAN discovery---finding agents within radio range or on the same local network---spans both underlays.

\paragraph{Outline.}
Section~\ref{sec:api} specifies the underlay-agnostic API the madGLP runtime calls.  Section~\ref{sec:proximity} specifies proximity PAN/LAN-based discovery and communication: its principles, then its realizations. Section~\ref{sec:ip} specifies the WAN realization.  Section~\ref{sec:system-predicates} specifies the layer functions behind the ten system predicates through which GLP agent code reaches the layer.  Section~\ref{sec:rendezvous} specifies the rendezvous server as a separate process, by its networking requirements only.  Section~\ref{sec:summary} summarizes the complete API and its correspondence to madGLP.

\section{Networking API}
\label{sec:api}
This section specifies the underlay-agnostic interface between the madGLP runtime and the networking layer.  The madGLP runtime issues these calls without regard to how a message travels; the PAN (Section~\ref{sec:ble}), LAN (Section~\ref{sec:lan-discovery}) and WAN (Section~\ref{sec:ip}) realizations carry them.

Throughout, a peer is named by its public key.  An \code{Underlay} is one of \code{PAN}, \code{LAN} or \code{WAN}; a peer may be reachable over more than one underlay at a time.  The \emph{proximity underlays} are \code{PAN} and \code{LAN}, on which a nearby agent is discovered (Section~\ref{sec:proximity}).  A LAN connection and a WAN connection are both IP connections; the layer tells them apart by whether the peer's address is on the agent's own local network (Section~\ref{sec:trust-levels}).  An IP connection is a \code{LAN} connection when the layer runs \code{LAN} and the peer's address is local; otherwise it is a \code{WAN} connection when the layer runs \code{WAN}; otherwise the layer neither makes nor accepts it.

The underlays are independent realizations: an agent may implement any subset of them and remains a conforming agent---identity, known peers, sessions, point-to-point delivery, trust, and the networking assumptions (Section~\ref{sec:assumptions}) are unchanged, and an unimplemented underlay is simply absent from discovery and \func{peerUnderlays}, and has no sub-object (Section~\ref{sec:instantiation}).  In particular, a server-side agent---a rendezvous server (Section~\ref{sec:rendezvous}), or a service co-hosted with one---realizes the WAN underlay alone, headless, with proximity discovery absent.

\subsection{Identity}
\label{sec:identity}
Each agent has an Ed25519 key pair.  The 32-byte public key is the agent's globally unique identity, used for addressing, authentication, and signing, and is persistent across sessions.  The layer holds the private key, signs with it and authenticates the sender of every message, and is part of the application the platform attests (Section~\ref{sec:session}).  The key pair is generated and persisted above the networking layer and is supplied to the layer's constructor when it is instantiated; it is immutable for the layer's lifetime.  Key rotation or a move to another device is realized by instantiating the layer with another pair.
\begin{longtable}{L{4cm} L{4.5cm} L{5.5cm}}
\toprule
\rowcolor{headerblue}
\textcolor{white}{\textbf{Function}} &
\textcolor{white}{\textbf{Signature}} &
\textcolor{white}{\textbf{Description}} \\
\midrule
\endhead
Retrieve identity & \func{getIdentity()} $\to$ \code{PubKey} & Return the agent's public key. \\
\bottomrule
\end{longtable}

\subsection{Instantiation}
\label{sec:instantiation}
The layer is created once per agent, with everything it needs to run; nothing in it changes for the layer's lifetime.
\begin{longtable}{L{4cm} L{4.5cm} L{5.5cm}}
\toprule
\rowcolor{headerblue}
\textcolor{white}{\textbf{Parameter}} &
\textcolor{white}{\textbf{Type}} &
\textcolor{white}{\textbf{Description}} \\
\midrule
\endhead
Key pair & Ed25519 key pair & The agent's identity (Section~\ref{sec:identity}). \\
\rowcolor{rowgray}
Underlays & \code{Underlay[]} & The underlays the layer runs: a subset of \code{PAN}, \code{LAN}, \code{WAN}; a headless server runs \code{WAN} alone (Section~\ref{sec:api}). \\
Server address & \code{Address?} & A rendezvous server's listening address (Section~\ref{sec:rendezvous}); null on a smartphone. \\
\rowcolor{rowgray}
Rendezvous server & boolean & Whether the agent is a rendezvous server (Section~\ref{sec:rendezvous}).\\
\bottomrule
\end{longtable}
The layer exposes one sub-object for each underlay it runs, \code{pan}, \code{lan} and \code{wan}, and null for an underlay it does not run.  The functions of this section are called on the layer itself, as they span all underlays: the layer chooses the underlay that carries each message, and reports reachability over all of them, so the callbacks that concern a peer carry the \code{Underlay} they refer to.  The functions that belong to one underlay are called on its sub-object, and take no underlay parameter: discovery and the trust level on \code{pan} and \code{lan} (Section~\ref{sec:proximity-principles}), and the public address, peer addresses, hole punching, the rendezvous servers and the local network identity on \code{wan} (Section~\ref{sec:ip}).  Thus, a call that makes no sense on an underlay, for example a trust level on the WAN, cannot be written, and a call on an underlay the layer does not run finds no sub-object.

\subsection{Connection and Reachability}
\label{sec:reachability}
The networking layer notifies the runtime when peers become reachable or unreachable, over any underlay, and reports the underlay(s) by which a peer is currently reachable.  Reporting the underlay lets GLP make proximity-aware decisions---for example, GLP computes which of its friends are physically nearby by intersecting its friends list with the peers currently reachable over \code{PAN}.  Which underlay carries a given message, and which path or rendezvous server is used, remain the networking layer's choice.  The connection report says whether the peer's identity is attested: it carries the attested application, or \code{null} for an unattested identity (Section~\ref{sec:session}), and the runtime reports it to the program~\cite{shapiro2026implementing}; what an unattested peer may be used for is GLP's decision.
\begin{longtable}{L{4cm} L{5cm} L{5cm}}
\toprule
\rowcolor{headerblue}
\textcolor{white}{\textbf{Function}} &
\textcolor{white}{\textbf{Signature}} &
\textcolor{white}{\textbf{Description}} \\
\midrule
\endhead
Peer connected & \func{onPeerConnected(pk, underlay, application?)} & Callback when a peer becomes reachable for communication.  Receives the peer's public key, the \code{Underlay} over which it connected, and, for an attested identity, the attested application: its developer, its name and its version number; \code{null} for an unattested identity (Section~\ref{sec:session}). \\
\rowcolor{rowgray}
Peer disconnected & \func{onPeerDisconnected(pk, underlay)} & Callback when a peer is no longer reachable over a given \code{Underlay}.  Receives the peer's public key and that underlay. \\
Check reachability & \func{isPeerReachable(pk)} $\to$ \code{bool} & Whether the given peer is currently reachable via any underlay. \\
\rowcolor{rowgray}
Reachable underlays & \func{peerUnderlays(pk)} $\to$ \code{List<Underlay>} & The underlays by which the given peer is currently reachable (empty if unreachable). \\
\bottomrule
\end{longtable}

\subsection{Known Peers}
\label{sec:known-peers}
The layer holds no social graph.  It holds a \emph{known-peer set}: the public keys GLP has supplied, each optionally carrying a last recorded address.  This set is what Closed trust (Section~\ref{sec:trust-levels}) recognizes, what unsolicited inbound WAN contact is checked against, what reconnection attempts range over, and where \func{putPeerAddress} (Section~\ref{sec:ip-connectivity}) records addresses.  Whether a known peer is a friend is GLP's interpretation; removing a key removes the layer's recognition of it.
\begin{longtable}{L{4cm} L{4.5cm} L{5.5cm}}
\toprule
\rowcolor{headerblue}
\textcolor{white}{\textbf{Function}} &
\textcolor{white}{\textbf{Signature}} &
\textcolor{white}{\textbf{Description}} \\
\midrule
\endhead
Add known peer & \func{putKnownPeer(pk)} & Add the public key to the known-peer set. \\
\rowcolor{rowgray}
Remove known peer & \func{removeKnownPeer(pk)} & Remove the key and any recorded address from the known-peer set. \\
\bottomrule
\end{longtable}

\subsection{ANNOUNCE and Liveness}
\label{sec:announce}
ANNOUNCE is the protocol's identity beacon and heartbeat.  It is a self-contained signed record whose payload carries:
\begin{itemize}
\item the sender's full 32-byte public key;
\item the protocol version (2 bytes);
\item a trailing Ed25519 signature over all preceding bytes.
\end{itemize}
The signature is verified against the carried sender key before the record is accepted, and repeated ANNOUNCEs from the same key are idempotent.

\paragraph{Role per underlay.}
ANNOUNCE serves a different mix of purposes on each underlay:
\begin{itemize}
\item \textbf{Over the PAN} it is, first, the \emph{identity bootstrap} of a first contact under Open trust: a scan match of an unrecognized presence carries no public key (Section~\ref{sec:ble-discovery}), so ANNOUNCE supplies the key that the Noise~XX handshake that follows checks (Section~\ref{sec:session}).  A recognized presence needs no ANNOUNCE before the handshake, as its token already names the key the IK handshake expects.  Second, it is a second \emph{liveness check}: link death is normally reported by the link layer itself, and the announce-driven sweep covers cases where no disconnect is ever surfaced, which would otherwise leave the peer attached indefinitely.  The staleness clock counts any authenticated traffic; ANNOUNCE guarantees a floor of traffic on an idle link.  When this check marks a connection as stale, the agent severs both GATT legs before synthesizing the disconnect.
\item \textbf{Over LAN and WAN} it is the \emph{liveness signal}: a UDP path has no connection state; a session silent for two consecutive intervals is torn down, feeding \func{onPeerDisconnected} (Section~\ref{sec:reachability}).  Over the WAN, identity plays no bootstrap role: a WAN connection is initiated by address with the peer's key already expected (Section~\ref{sec:session}); on the LAN, ANNOUNCE bootstraps identity only for a first contact under Open trust, as on the PAN.  An inbound connection is identified by its completed handshake.
\end{itemize}

\paragraph{Heartbeat.}
Once the session is established, ANNOUNCE is sent inside it, and repeats periodically (every 10\,s in the implementation) on every live connection of every underlay; only the ANNOUNCE of a first contact under Open trust travels before the handshake. A peer silent for two consecutive intervals is stale \emph{on that underlay}, independently per underlay, with the per underlay consequences above. The same heartbeat tick drives reconnection attempts towards unreachable known peers and rendezvous servers and the replay of queued messages once a recipient is live again (fair delivery, Section~\ref{sec:assumptions}).

\paragraph{Trust gating.}
Under Closed trust on a proximity underlay (Section~\ref{sec:trust-levels}), no ANNOUNCE is sent or answered before the handshake: the agent contacts a recognized peer with the IK handshake, and an ANNOUNCE arriving before a handshake is ignored and its connection dropped.  Unsolicited WAN contact is gated by the known-peer set instead (Section~\ref{sec:trust-levels}). Address distribution is not ANNOUNCE's role: peer addresses reach the layer via \func{putPeerAddress} (Section~\ref{sec:ip-connectivity}), fed by GLP.  Over LAN and WAN every connected peer is already authenticated.

\subsection{Session Establishment}
\label{sec:session}
Once a transport-level connection to a peer exists---a GATT connection on the PAN (Section~\ref{sec:ble}) or a UDP path on the LAN or WAN (Section~\ref{sec:ip})---the peers perform a mutual authentication handshake using their Ed25519 keys via Noise (Curve25519 for key exchange, ChaCha20-Poly1305 for encryption, SHA-256 for hashing).  The handshake proves that each peer holds the private key of its static public key, and derives forward-secret session keys.  The Noise pattern depends on what the initiator already knows.  When the initiator expects a specific responder key, it runs Noise~IK: it sends its own static key in the first message, encrypted to the expected key, so only the holder of that key can read or answer it, and the responder never transmits its static key.  The initiator expects a key for a known peer (Section~\ref{sec:known-peers}), for a recognized proximity token (Section~\ref{sec:proximity-principles}), for a rendezvous server, whose key GLP declared (Section~\ref{sec:ip-connectivity}), and on every WAN connection, since the layer connects over the WAN only to known peers (Section~\ref{sec:ip-cold-call}).  Only a first contact with an unrecognized presence under Open trust (Section~\ref{sec:trust-levels}) has no expected key: there the peers exchange ANNOUNCE first (Section~\ref{sec:announce}) and run Noise~XX, in which each side transmits its static key, and each verifies the delivered key against the key ANNOUNCE claimed; a mismatch aborts the connection.  The responder enforces the trust level: under Closed trust, and on the WAN except on a rendezvous server, it answers an IK first message only from a key in the known-peer set, and answers no XX first message, so a stranger never learns its key.  A recorded IK first message can be replayed, and only the responder can answer it, so its recorder can test whether the same responder is still present, though it learns no key.  The Curve25519 static key carried by Noise is the birational image of the agent's Ed25519 identity key, so the expected static is computable from the peer's Ed25519 public key alone; no separate key distribution is required.  The handshake needs no ordered carrier: it is a strict request--response exchange, self-ordering by construction: each message is sent only after the previous one arrived; over a lossy carrier each handshake message is retransmitted on a timer until the next one arrives, and the final one, which has no reply (message~2 of IK, message~3 of XX), until the other side's attestation message arrives (below), transport messages are sent only after the handshake completes, and the session's replay protection makes duplicated packets harmless.  Sessions are per underlay: one handshake per underlay per pair, so a pair reachable over more than one underlay holds a session on each. A peer becomes reachable over an underlay (Section~\ref{sec:reachability}) once its session on that underlay is established and authenticated, and the attestation exchange on it has completed, with the peer either attested or declared unattested.

\paragraph{Attestation.}
The handshake is followed by a mutual attestation exchange, and neither peer is reported reachable until it completes.  The exchange has two parts, one long-lived and one per session.  Once, the layer asks the platform to generate an \emph{attestation key} in its secure element and to attest it over a challenge naming the agent's identity public key $pk$: App Attest on iOS, hardware key attestation on Android.  For each session, the layer asks the platform to sign with that key the digest $H(\text{``glp attest''} \mid pk \mid v \mid h)$, where $v$ is the application's version number and $h$ is the final Noise handshake hash.  Each side sends one attestation message per session, as a message of the Message Protocol (Section~\ref{sec:message-transport}) after the final handshake message: either its attestation and this signature, or a statement that its platform provides no attestation.  The receiver verifies the attestation against the platform's root, and the signature against the key the attestation carries; either failing tears the session down, while the identity of a peer that sends the statement is reported unattested: \func{onPeerConnected} carries \code{null} for the application.  Since any agent can declare itself unattested, including one running a repackaged application or holding a copied identity key, attestation protects only the uses GLP restricts to attested peers.  The receiver waits for the peer's attestation message as long as the sender's attempts to deliver it last, then tears the session down; an attestation message is never queued for the next session.  A payload arriving on a session whose exchange has not completed is dropped unacknowledged, so the sender's retransmission delivers it once the exchange has completed.

The attestation binds the attestation key to the application and to the identity key, and the signature binds both, with the version number, to this session.  We bind the signature to the handshake hash because an identity key is long-lived: an attestation over the key alone could be replayed onto every session the agent opens, while $h$ is unique per session, identical at both ends, and chosen by neither side alone.  The attestation itself is long-lived, since a platform fixes the challenge when it generates the key, and a per-session challenge would need a new secure-element key per session.  On iOS the attestation key remains valid through updates of the application, so an updated application needs no new attestation and signs its new version number in each new session; reinstalling the application, migrating it to another device or restoring the device from a backup needs a new attestation key~\cite{appleAppIntegrity}.  Nothing of the session is cached, so a peer that restarts proves itself again before it is reachable.  The operating system above the runtime relies on two properties of this exchange: an attestation cannot be replayed onto another session, and the layer only verifies attestations, while GLP decides what an unattested peer, such as a rendezvous server (Section~\ref{sec:rendezvous}), may be used for.

The key the platform holds is the attestation key, not the identity key.  The identity key is Ed25519 with the Curve25519 static above, and hardware key stores hold other curves---Apple's Secure Enclave holds P-256 only---so the identity key lives in the application's own store.  The attestation key is generated in the secure element and never leaves it; the attestation binds the identity key to it once, and the signature binds it to each session.  A copy of an identity key is therefore worth less, but not nothing: on a machine that does not run the attested application it opens only unattested sessions, and a genuine application that imports keys, for example to move an identity to a new smartphone~\cite{eitan2026securing}, attests it like any key it generated.  Attestation proves that a genuine application holds the key, not that the key's owner does.

The two platforms do not prove the same thing, and the difference bounds the claim above.  Both attestations are verified by the peer against a public root---Apple's App Attest root, Google's hardware attestation root---so no third party is contacted at the moment two agents establish a session, which is the property the protocol needs.  Google's Play Integrity is not used: its verdict is opened only by Google's servers, so proving oneself to a peer would require reaching a global service on every session.  What Android's certificate chain establishes is that the attestation key lives in the device's secure hardware and that the application holding it carries the developer's signing certificate the attestation names; on a device whose operating system is modified that certificate can be forged.  The claim above therefore holds against a repackaged application on both platforms, and against a modified operating system on neither.

Only the device's hardware root can prove which application the device runs, and the platform's vendor holds that root; neither code the application ships nor a server can prove it, since each sees only what the device sends.  What the platforms do require of a third party lies outside the session, and the two differ.  On Android the attestation certificates are provisioned to the device by Google and, from Android~16, only so, and revocation is published by Google as a status list a verifier fetches; so a device that has never reached Google cannot attest, and a verifier that never reaches it does not learn that a key has been revoked.  On iOS the corresponding contact is on the producing device alone: attesting a key reaches an Apple server once, and Apple documents no revocation, so an iOS verifier contacts nobody at all.  Neither platform is contacted during a session, and an agent that cannot attest is reported unattested (below) rather than refused, so a session between two agents remains a matter for the two of them.

What the attestation names is the application published under its developer's key, not the code.  To report the same on both platforms, we take the attested application to be its developer, its name and its version number.  Android's certificate names the digest of the developer's signing certificate, the package and its version; App Attest names the application identifier the key was attested under, the developer's team identifier and the bundle identifier, and no version.  On both platforms the version number is the one the application signs in the per-session digest, which only the genuine application can sign.  Neither platform measures the running code, so that one build runs under one name and version number is the developer's undertaking, which the operating system above assumes.

A headless server profile (Section~\ref{sec:api}) has no attestation to offer: it sends the statement that its platform provides no attestation, and its peers report it unattested rather than refusing it; whether to transact with an unattested peer is a GLP-level decision.  An attestation offered and found invalid tears the session down.

\subsection{Geofencing}
\label{sec:geofencing}
The location service is the platform's geofencing, and the layer uses it for one purpose: to tell GLP when the smartphone enters or leaves a place GLP has declared.  A place is a geofence, a circle of a given radius in metres about the location the device is at when the place is declared, registered with the platform under a name GLP chooses.  With \func{declarePlace(place, radius)} the layer registers the geofence, and a second declaration of the same place replaces the first; with \func{removePlace(place)} it unregisters it (Section~\ref{sec:system-predicates}).  The platform then watches the device's position and reports each crossing of a registered geofence to the layer, which passes it on through \func{onPlaceEvent(place, event)} as \code{entered} or \code{exited}.  The event carries the name of the place and the direction of the crossing, and no coordinates: nothing of the device's location reaches the runtime or a peer through the layer.  What a place means, for example that the person declared the place they are in trusted, so PAN contact there is open (Section~\ref{sec:trust-levels}), is decided by GLP, and the layer makes no decision about it.

The platform does not always report crossings: it stops when the application moves to the background or when the person withdraws the location permission, and resumes when either is reversed.  Without knowing this, GLP could not tell a device that stayed inside a place from one whose exit went unreported, so the layer reports \code{unobservable} when the platform stops reporting and \code{observable} when it resumes.  Observability belongs to the platform's reporting and not to one geofence, so both events are delivered for every standing place at once, and a place declared while reporting is stopped is told so as soon as it is declared.  A crossing that occurs while the device is unobservable is lost, and on resumption the layer replays nothing.  On iOS, the platform reports crossings to an application in the background only under the permission to use location at all times, which we do not request, so the platform supplies the position only in the foreground: the layer decides the crossing itself, and reports the device unobservable whenever the application is not in the foreground.  A headless server has no location service: its layer is built without one, and refuses \func{declarePlace} and \func{removePlace}.

\subsection{Cold Calls}
\label{sec:cold-call}
A \emph{cold call} is a first-contact connection attempt between two agents that do not already share an active communication channel.  A cold call is realized in one of three ways: by an encounter on the PAN (Section~\ref{sec:ble-discovery}), by a LAN encounter (Section~\ref{sec:lan-discovery}), or over the WAN between agents whose programs have declared each other's keys (Section~\ref{sec:ip-cold-call}). What a cold call \emph{means} is a GLP-level interpretation and is not part of the networking layer.

\paragraph{Trust levels.}
\label{sec:trust-levels}
Whether the networking layer answers an \emph{unsolicited} proximity encounter from an unknown key (Section~\ref{sec:proximity-principles}) is governed by a trust level set by GLP and enforced by the layer:
\begin{itemize}
\item \textbf{Open:} The agent completes the ANNOUNCE exchange and the Noise~XX handshake with any unknown contact (Section~\ref{sec:session}), thereby revealing its full public key and allowing first contact.
\item \textbf{Closed:} The agent sends no ANNOUNCE, answers no XX handshake, and answers an IK handshake only from a known key: unmatched proximity PAN and LAN encounters are ignored, and the agent reveals its key only to peers that already hold it.
\end{itemize}
A level is held per proximity underlay, and is set with \func{setTrustLevel(level)} on that underlay's sub-object, \code{pan} or \code{lan} (Section~\ref{sec:instantiation}); both levels default to Closed.  WAN contact is governed by no level (below). GLP sets the two levels independently, for example opening PAN contact while the smartphone is inside a place the person declared trusted (Section~\ref{sec:geofencing}), and LAN contact while it is attached to a local network the person declared trusted (Section~\ref{sec:network-identity}).

Unsolicited inbound WAN contact from outside the agent's own local network is not governed by trust levels: it is refused unless the peer is in the known-peer set (Section~\ref{sec:known-peers}): an IK handshake from an unknown key is not answered (Section~\ref{sec:session}).  A rendezvous server, marked so when its layer is instantiated (Section~\ref{sec:instantiation}), answers unsolicited WAN contact from any key, since it cannot know its clients in advance (Section~\ref{sec:rendezvous}).

LAN contact arrives on the same path, LAN and WAN connections being both IP connections, so the layer tells the two apart by the source address alone: an unsolicited inbound contact whose source address lies within one of the agent's own local address prefixes is LAN contact, governed by the LAN level.
\subsection{Point-to-Point Communication}
\label{sec:p2p}
The networking layer provides reliable delivery of arbitrary byte payloads between peers, identified by public key.
\begin{longtable}{L{4cm} L{5.5cm} L{4.5cm}}
\toprule
\rowcolor{headerblue}
\textcolor{white}{\textbf{Function}} &
\textcolor{white}{\textbf{Signature}} &
\textcolor{white}{\textbf{Description}} \\
\midrule
\endhead
Send payload & \func{send(pubKey, payload)} & Send a byte payload to the peer identified by \code{pubKey}. \\
\rowcolor{rowgray}
Receive payload & \func{onMessageReceived(sender, payload, messageId, underlay)} & Register a callback for incoming payloads.  Receives the sender's public key, the byte payload, a layer-assigned \code{messageId}, and the \code{Underlay} over which the message was delivered. \\
\bottomrule
\end{longtable}

\subsection{Message Protocol}
\label{sec:message-transport}
One message protocol carries \func{send} payloads, and the attestation exchange of Section~\ref{sec:session}, on every underlay. Every packet bears the five-byte frame: a packet-type byte and a 4-byte big-endian payload length.  A message is assigned its \code{messageId} and split into fragments sized to the carrier, each fragment carrying the \code{messageId}, its index, and the fragment count, encrypted under the session of Section~\ref{sec:session}.  The receiver acknowledges fragments; the sender retransmits unacknowledged fragments with backoff, under a bounded send window per peer. When all fragments have arrived the message is reassembled and delivered whole---atomic delivery (Section~\ref{sec:assumptions}).  A re-send after a lost acknowledgment, possibly over another underlay, may deliver a message multiple times, and madGLP dedupes repeated deliveries~\cite{shapiro2026implementing}. When a message's attempts are exhausted the session is torn down and the message queues for the next path (Section~\ref{sec:announce}). No order is kept or promised across messages (Section~\ref{sec:assumptions}); a lost fragment delays only its own message.

Over LAN and WAN, a fragment rides a UDP datagram, sized within the path MTU, and retransmission is the reliability mechanism; over BLE, a fragment rides a GATT write without response, sized to the negotiated MTU, and retransmission is the reliability mechanism on both underlays.

\subsection{Networking Assumptions}
\label{sec:assumptions}
The networking layer must satisfy the following assumptions for madGLP correctness:

\noindent\textbf{Fair message delivery:} Every message passed to \func{send()} is eventually delivered to the recipient at least once, assuming the recipient eventually becomes reachable.  Messages to temporarily-unreachable peers are queued and delivered when connectivity is restored; the queue survives the termination of the application.

\noindent\textbf{Unordered delivery:} No delivery-order assumption is made: messages between two agents may be delivered in any order, and the madGLP runtime tolerates arbitrary delivery order.  The networking layer is free to route concurrent messages to the same peer over different underlays.

\noindent\textbf{Atomic message delivery:} Each message is delivered as a complete unit.  If a message is fragmented across multiple packets, the networking layer reassembles them before delivery.

\noindent\textbf{Integrity:} Messages are not corrupted in transit.  The networking layer provides encryption and integrity checks.

\noindent\textbf{Authentication:} The sender's identity (public key) is authenticated.  A received message from public key $pk$ was indeed sent by the agent holding the corresponding private key.

\bigskip

\section{Proximity Underlays: PAN and LAN}
\label{sec:proximity}
An agent discovers and contacts agents in its vicinity---on the same local network, or within radio range---without any infrastructure and without becoming trackable. This section specifies the shared design, then its two realizations: PAN and LAN.

\subsection{Concepts and Principles}
\label{sec:proximity-principles}
\paragraph{Recognition before contact.}
Each proximity underlay gives an agent a public, passively observable presence---a BLE advertisement, an mDNS service instance.  That presence must let a peer who already knows the agent recognize it before any contact, yet give everyone else nothing stable to track.  Both realizations therefore advertise the same construction: a rotating 8-byte \emph{recognition token}, the first 8 bytes of $H(L \mid pk \mid T)$, where $H$ is SHA-256, $pk$ the advertiser's public key, $T$ the current 15-minute time slot, and $L$ a per-underlay label. Anyone who knows $pk$ computes the token for the current and adjacent slots and matches it against observed presences; adjacent-slot matching absorbs clock skew. To anyone else the token is an identifier that changes every slot. The advertisement carries the token and no public key.
\paragraph{Contact and identity.}
A recognized presence is contacted on its underlay with the IK handshake of Section~\ref{sec:session}, whose expected key is the recognized peer's.  Under Open trust any other Grassroots presence is contacted too: the peers exchange ANNOUNCE (Section~\ref{sec:announce}), binding the contact to a claimed public key, and run the XX handshake, which checks the claim.  Where either endpoint may initiate the connection, the deterministic initiator is the endpoint with the lexicographically smaller public key, as everywhere (Section~\ref{sec:ip-connection}).  Addresses and contacts observed by proximity discovery feed the layer's connection attempts directly; GLP distributes addresses (Section~\ref{sec:ip-connectivity}).

\paragraph{Cold-call trust levels.}
The trust levels of Section~\ref{sec:trust-levels} govern proximity encounters as unsolicited contact, one level per proximity underlay: under Open the agent contacts or answers unmatched presences on that underlay, revealing its full public key in ANNOUNCE; under Closed the advertised token serves only to recognize already-known peers (Section~\ref{sec:known-peers}) and unmatched encounters are ignored.  The PAN and LAN underlays carry independent trust levels, so an agent may answer strangers over one underlay while ignoring them over the other.

\paragraph{Discovery.}
\label{sec:ble-discovery-api}
Proximity discovery is exposed to the runtime by the same functions on both proximity sub-objects, \code{pan} and \code{lan} (Section~\ref{sec:instantiation}), each acting on its own underlay:
\begin{longtable}{L{4cm} L{4.5cm} L{5.5cm}}
\toprule
\rowcolor{headerblue}
\textcolor{white}{\textbf{Function}} &
\textcolor{white}{\textbf{Signature}} &
\textcolor{white}{\textbf{Description}} \\
\midrule
\endhead
Scan for peers & \func{getPeers()} $\to$ \code{List<Peer>} & Look for nearby Grassroots agents on this underlay: on \code{pan} a BLE scan for the Grassroots prefix, on \code{lan} an mDNS browse for the service type. \\
\rowcolor{rowgray}
Peer discovered & \func{onPeerDiscovered(pk)} & Callback for each nearby Grassroots agent found on this underlay. \\
Set trust level & \func{setTrustLevel(level)} & Set the cold-call trust level of this underlay to Open or Closed (Section~\ref{sec:trust-levels}).\\
\bottomrule
\end{longtable}

\subsection{LAN}
\label{sec:lan-discovery}
Two agents on the same local network can find each other and communicate directly---no BLE radio and no rendezvous server.  Discovery uses DNS-SD over multicast DNS: each agent advertises an instance of the service type \code{\_grassroots.\_udp} on the local domain and browses for instances of it.  The advertised instance name is the recognition token of Section~\ref{sec:proximity-principles}, hex-encoded, with label \code{"glp lan suffix"}; the instance's SRV record carries the agent's LAN address and bound UDP port.

A matched instance yields an address, and contact proceeds as over the WAN: a UDP path with the deterministic initiator (Section~\ref{sec:ip-connection}) and the session of Section~\ref{sec:session}, preceded by ANNOUNCE only for an unrecognized instance under Open trust; reachability reports the underlay \code{LAN}, and communication is that of Section~\ref{sec:ip-comm}.

\subsection{PAN}
\label{sec:ble}
Bluetooth Low Energy enables offline, infrastructure-free operation between devices in direct radio range of each other, and is the bootstrap underlay where agents first meet face-to-face.

\subsubsection{Advertising and Recognition}
\label{sec:ble-identity}
Each agent advertises a BLE GATT service whose 128-bit (16-byte) UUID has two parts: a fixed 8-byte \emph{Grassroots prefix} common to all Grassroots agents, enabling general discovery of any nearby Grassroots agent, and the rotating 8-byte recognition token of Section~\ref{sec:proximity-principles} with label \code{"glp ble suffix"}. When the slot time is over, the API rotates the new token, and advertising is restarted with the new discovery token. When an app is not in the foreground, iOS advertises the service only to a scanner for that exact UUID, so a friend who computes the token still finds the agent; discovery by strangers needs the app in the foreground.

\subsubsection{Connection and Identity}
\label{sec:ble-discovery}
Upon BLE connection to a recognized presence, the peers establish the IK session of Section~\ref{sec:session} over the GATT connection directly; to an unrecognized presence under Open trust, they first exchange ANNOUNCE, which carries the agent's full 32-byte public key, and then run the XX handshake.  No Bluetooth bonding or pairing confirmation is required.  Where either endpoint may initiate---both run central and peripheral---the deterministic initiator is the endpoint with the lexicographically smaller public key, as over the WAN (Section~\ref{sec:ip-connection}).

\subsubsection{Communication}
\label{sec:ble-comm}
\func{send(pubKey, payload)} transmits the byte payload over the established GATT connection via the message protocol of Section~\ref{sec:message-transport}, GATT writes as the carrier; fragmentation and reassembly satisfy atomic delivery (Section~\ref{sec:assumptions}).  \func{onMessageReceived} delivers reassembled payloads to the runtime.

\bigskip

\section{WAN Underlay}
\label{sec:ip}
This section specifies how the networking API (Section~\ref{sec:api}) is realized over the WAN, the Internet at large. Communication uses datagrams carrying the message protocol of Section~\ref{sec:message-transport} over UDP sockets. NAT traversal is coordinated by a rendezvous agent, on a server or on a well-connected friend's smartphone (Section~\ref{sec:rendezvous}). The rendezvous protocol is GLP code (Section~\ref{sec:rendezvous}).

\subsection{WAN Connection}
\label{sec:ip-connection}
Once a UDP path is established (directly or by hole-punching), the peers establish the session of Section~\ref{sec:session} over it. When both endpoints might initiate---mutual LAN discovery, simultaneous reconnection---the deterministic initiator is the endpoint with the lexicographically smaller public key. UDP has no connections, so ANNOUNCE is the liveness check (Section~\ref{sec:announce}): a peer silent for two intervals is disconnected.

\subsection{Connectivity and Address}
\label{sec:ip-connectivity}
Mobile agents typically run behind NAT without static IP addresses, and an agent cannot directly observe its own public address.  When its address changes (e.g., due to roaming), all its IP-based paths break. The networking layer reports address changes to the runtime, and exposes the agent's current public address once known (an agent learns it from a rendezvous server, which observes the source address of the agent's packets; see Section~\ref{sec:rendezvous}).  GLP code uses these to share the agent's address with its peers and to trigger reconnection.  The address is GLP data and travels over any underlay: GLP can hand a fresh address to a BLE-adjacent friend, which can relay it to remote friends and coordinate their reconnection via the system predicates (Section~\ref{sec:system-predicates}), serving as an edge rendezvous point without a globally-routable address.

\paragraph{Address-change recovery.}
When an agent's own address changes, its established UDP paths break and their sessions are torn down by liveness (Section~\ref{sec:announce}), but its ability to initiate outbound connections is unimpaired: the recorded addresses of its known peers did not change.  The heartbeat sweep therefore re-initiates connections to the known peers and the declared rendezvous servers; the rendezvous connection always succeeds (globally-routable target), restoring the RV channel and, through it, the agent's fresh public address.  Recovery is thus asymmetric by design: the endpoint that moved reconnects; a stationary peer is reconnected to, and needs no action beyond accepting.  Distributing the fresh address to peers that cannot be reached directly (both behind NAT) is GLP's: GLP learns the address from \func{onConnectivityStatusChanged}/\func{getPublicAddress}, sends it to friends over whatever channels remain live---rendezvous-mediated signaling, a BLE-adjacent friend, surviving sessions---and it enters each peer's layer via \func{putPeerAddress}, whereupon that peer's own sweep connects to it, punching if coordinated (Section~\ref{sec:system-predicates}).

\paragraph{Rendezvous servers.}
The rendezvous servers the layer connects to are those GLP has declared, each by its public key and its address, and GLP adds and removes them while the layer runs.  GLP knows which of its peers are rendezvous servers by the keys it declared.

The functions below are called on the layer's \code{wan} sub-object (Section~\ref{sec:instantiation}).
\begin{longtable}{L{3.4cm} L{6.2cm} L{5cm}}
\toprule
\rowcolor{headerblue}
\textcolor{white}{\textbf{Function}} &
\textcolor{white}{\textbf{Signature}} &
\textcolor{white}{\textbf{Description}} \\
\midrule
\endhead
Connectivity changed & \func{onConnectivityStatusChanged\allowbreak{}(oldAddress?, newAddress?)} & Callback when the networking layer detects a change in the agent's public IP address. Receives \code{(oldAddress?, newAddress?)} as \code{ip:port} strings, either of which may be \code{null} when no public address is available (e.g. startup: \code{old=null}; loss: \code{new=null}; update: both present and different).  Fired on startup and on every address change. \\
\rowcolor{rowgray}
Public address & \func{getPublicAddress()} $\to$ \code{Address?} & The agent's current public \code{ip:port}, or \code{null} if not yet known. \\
Feed peer address & \func{putPeerAddress(pk, address)} & Inform the networking layer of a peer's current public \code{ip:port}, learned at the GLP level; the layer records it with the known peer (Section~\ref{sec:known-peers}) for \func{send} and reconnection. One WAN address is recorded per peer; the latest, whether GLP supplied it or the layer observed it, replaces the earlier.\\
\rowcolor{rowgray}
Observed address & \func{observedPeerAddress(pk)} $\to$ \code{Address?} & The public \code{ip:port} at which the layer observes the connected peer, or \code{null} if none is observed; behind \code{peer\_address/2}, which reports \code{address(S)} or \code{none}~\cite{shapiro2025glp}. \\
Punch & \func{punchUdp(address)} & Send toward the address a short burst of UDP datagrams to open the NAT mapping; behind \code{punch\_udp/1}. \\
\rowcolor{rowgray}
Declare server & \func{putRendezvousServer(pk, address)} & Add the rendezvous server with public key \code{pk} at \code{address} to the servers the layer connects to; a further declaration of the same \code{pk} replaces its address; behind \code{rendezvous\_declare/2}. \\
Remove server & \func{removeRendezvousServer(pk)} & Remove the server from those the layer connects to, and end the layer's session with it; a key not declared does nothing; behind \code{rendezvous\_remove/1}. \\
\bottomrule
\end{longtable}

\subsection{Local Network Identity}
\label{sec:network-identity}
The \code{wan} sub-object (Section~\ref{sec:instantiation}) reports an opaque fingerprint of the local network the agent is attached to, and notifies the runtime when it changes.  The fingerprint is derived from whichever of these the platform provides without additional permission: the local address prefix, the default gateway address, and the wireless network name.

This is not \func{onConnectivityStatusChanged} (Section~\ref{sec:ip-connectivity}).  The two are independent in both directions: the attached local network can change while the public address does not, and the public address can change while the attached local network does not.

The fingerprint is best-effort: two distinct local networks may yield the same one, and a platform may make none of its constituents readable.  A match is evidence that the agent is attached to the same local network as before, not proof of it; an agent compares a fingerprint only against its own earlier values. 
\begin{longtable}{L{3.4cm} L{6.2cm} L{5cm}}
\toprule
\rowcolor{headerblue}
\textcolor{white}{\textbf{Function}} &
\textcolor{white}{\textbf{Signature}} &
\textcolor{white}{\textbf{Description}} \\
\midrule
\endhead
Network identity & \func{networkIdentity()} $\to$ \code{NetworkId?} & An opaque fingerprint of the local network the agent is currently attached to, or \code{null} when it is attached to none or no constituent is readable. \\
\rowcolor{rowgray}
Network changed & \func{onNetworkChanged(oldId?, newId?)} & Callback when the fingerprint changes. Receives \code{(oldId?, newId?)}, either of which may be \code{null}. \\
\bottomrule
\end{longtable}

\subsection{WAN Cold-Call}
\label{sec:ip-cold-call}
Over the WAN, a cold call is a connection between two agents whose programs have each declared the other's key to their layer (Section~\ref{sec:known-peers}).  How the two learn each other's keys---by a signed one-time invitation link redeemed at a rendezvous server---belongs to the operating system above the runtime (the Grassroots Super-App, in preparation) and is out of scope here.  To the layer each is a known peer of the other: the two connect over UDP, punching where needed (Section~\ref{sec:ip-connectivity}), and establish the session of Section~\ref{sec:session}; only then does the caller send the madGLP cold-call payload to the callee's index-0 serializer.

\subsection{WAN Communication}
\label{sec:ip-comm}
\func{send(pubKey, payload)} transmits the byte payload to the peer as session-encrypted fragments in UDP datagrams, per the message protocol of Section~\ref{sec:message-transport}.  \func{onMessageReceived} delivers complete, reassembled messages to the runtime.
\bigskip

\section{The Seam: Layer Functions behind the System Predicates}
\label{sec:system-predicates}
The GLP runtime realizes ten system predicates over the networking-layer functions this section specifies. These predicates give GLP agents the abilities of address observation, UDP hole punching, declaration of places as geofences, trust level setting, known-peer set management including agent addresses, and declaration of rendezvous servers.  The predicates---\code{peer\_address/2}, \code{punch\_udp/1}, \code{place\_declare/3}, \code{place\_remove/1}, \code{trust\_declare/2} with the values \code{pan} and \code{lan}, \code{peer\_declare/1}, \code{peer\_remove/1}, \code{address\_declare/2}, \code{rendezvous\_declare/2} and \code{rendezvous\_remove/1}---are defined in the language paper's catalogue~\cite{shapiro2025glp}, and their clauses in the implementation paper~\cite{shapiro2026implementing}; only the functions behind them are specified here.

If an underlay changes---for example, UDP datagrams replaced by QUIC, or hole-punching by another traversal mechanism---only the layer's implementation of these functions changes; the GLP code that uses the predicates is unaffected.

\paragraph{The layer functions.}
The functions behind the predicates are specified with their underlay or subject: \func{observedPeerAddress} and \func{punchUdp} with the WAN, on the \code{wan} sub-object (Section~\ref{sec:ip-connectivity}), \func{setTrustLevel} with the trust levels, on the \code{pan} and \code{lan} sub-objects (Section~\ref{sec:trust-levels}), \func{putKnownPeer} and \func{removeKnownPeer} with the known peers, on the layer (Section~\ref{sec:known-peers}), and \func{putPeerAddress}, \func{putRendezvousServer} and \func{removeRendezvousServer} with the WAN, on the \code{wan} sub-object (Section~\ref{sec:ip-connectivity}).  The place functions are specified here:
\begin{longtable}{L{4cm} L{4.5cm} L{5.5cm}}
\toprule
\rowcolor{headerblue}
\textcolor{white}{\textbf{Function}} &
\textcolor{white}{\textbf{Signature}} &
\textcolor{white}{\textbf{Description}} \\
\midrule
\endhead
Declare place & \func{declarePlace(place, radius)} & Register with the platform a geofence of \code{radius} metres about the device's current location, named \code{place}; a further declaration of the same \code{place} replaces the first; behind \code{place\_declare/3}. \\
\rowcolor{rowgray}
Remove place & \func{removePlace(place)} & Unregister the geofence of \code{place}; a \code{place} not declared does nothing; behind \code{place\_remove/1}. \\
Place events & \func{onPlaceEvent(place, event)} & Callback with the \code{place} and an \code{event}, one of \code{entered} and \code{exited}, as the platform reports the device crossing the geofence, \code{unobservable} when the platform stops reporting---the application in the background, a permission withdrawn---and \code{observable} when it resumes; feeds the stream of \code{place\_declare/3}. \\
\bottomrule
\end{longtable}

\paragraph{Signatures.}
The layer provides \code{sign(bytes)} and \code{verify(pk, bytes, signature)}, Ed25519 signing and verification under the agent's key and against a public key.  The runtime realizes the language's \code{sign/3} and \code{signature/2} over them~\cite{shapiro2025glp}, adding the source identity of the signing program, so the layer never learns about programs, as it never learns about ``friend''.

\bigskip

\section{Rendezvous Server}
\label{sec:rendezvous}
A \emph{rendezvous server} is a separate process---a host that provides a static, globally-routable IP address and runs a lightweight GLP \emph{rendezvous agent}.  Its purpose is to give NAT-bound agents, which cannot otherwise reach one another, a fixed point they can always reach outbound, from which they discover their own public addresses and coordinate UDP hole-punching.

This section specifies the rendezvous server only by its networking requirements---what a machine must provide to host one.  The rendezvous agent, and the reconnection and first-contact protocols that use it, belong to the operating system above the runtime (the Grassroots Super-App, in preparation) and are out of scope here.

\paragraph{Networking requirements.}
\begin{itemize}
\item \textbf{Static, globally-routable address.}  The host has a fixed public \code{ip:port} that any agent can reach outbound, traversing NAT in the outbound direction without assistance.
\item \textbf{Address observation.}  When an agent connects, the rendezvous server observes the source \code{ip:port} of the agent's packets and exposes it via \func{observedPeerAddress} (Section~\ref{sec:ip-connectivity}).  This is the address-discovery mechanism that lets a NAT-bound agent learn its own public address; no external STUN infrastructure is required.
\item \textbf{Hole-punch coordination.}  The rendezvous server relays observed addresses between two agents, and each sends punch traffic toward the other with \func{punchUdp} (Section~\ref{sec:ip-connectivity}) so a direct UDP path can form.  Once the path is open the peers run the Noise handshake and the message protocol directly (Sections~\ref{sec:ip-connection} and~\ref{sec:session}); the rendezvous server sees signaling metadata and observed addresses, never the encrypted payload.
\item \textbf{No relay.}  The rendezvous server is not a TURN relay; it never forwards madGLP payloads.  When direct connectivity cannot be established (e.g., symmetric NAT), relaying is an application-level concern handled in GLP by a connected common peer, not by the networking layer.
\item \textbf{Any key.}  A rendezvous server cannot know its clients in advance, so its layer, marked so when it is instantiated (Section~\ref{sec:instantiation}), answers unsolicited WAN contact from any key (Section~\ref{sec:trust-levels}); it belongs to no social graph.
\item \textbf{Unattested.}  Having no platform attestation to offer (Section~\ref{sec:session}), a rendezvous server is reported unattested, and this costs nothing: a false address fails to connect, and it cannot make an unattested peer appear attested, attestation being verified between the two agents themselves.  GLP knows a rendezvous server by the key it declared (Section~\ref{sec:ip-connectivity}).  What an unattested channel may then be used for is a GLP-level decision.
\end{itemize}

\paragraph{Any number of servers.} Any individual or organization can run a rendezvous server; the servers do not communicate with one another.  An agent may use several, which GLP declares (Section~\ref{sec:ip-connectivity}); the system operates with none when agents are on the same PAN or LAN (Section~\ref{sec:proximity}).  In the terminology of the ICE framework~\cite{rfc8445}, a rendezvous server combines the roles of STUN server (address discovery via \func{observedPeerAddress}) and application-specific signaling server (candidate exchange and hole-punch coordination, performed by the rendezvous agent); it does not serve as a TURN relay.

\section{Complete API Summary}
\label{sec:summary}
The following table summarizes the complete networking API.  The first five groups are underlay-agnostic and called on the layer; the Proximity group is called on the \code{pan} and \code{lan} sub-objects, and the WAN group on the \code{wan} sub-object (Section~\ref{sec:instantiation}); the system predicates are the language's~\cite{shapiro2025glp}, and the layer functions behind them are in Section~\ref{sec:system-predicates}.
\begin{longtable}{L{2.4cm} L{6.2cm} L{2.5cm} L{3.5cm}}
\toprule
\rowcolor{headerblue}
\textcolor{white}{\textbf{Category}} &
\textcolor{white}{\textbf{Function / Predicate}} &
\textcolor{white}{\textbf{Direction}} &
\textcolor{white}{\textbf{madGLP Role}} \\
\midrule
\endhead
\multicolumn{4}{l}{\textbf{Identity}} \\
\midrule
\rowcolor{rowgray}
Query & \func{getIdentity()} & Runtime $\leftarrow$ Network & Retrieve agent public key \\
\midrule
\multicolumn{4}{l}{\textbf{Known peers}} \\
\midrule
Add & \func{putKnownPeer(pk)} & Runtime $\to$ Network & Supply a key the layer recognizes \\
\rowcolor{rowgray}
Remove & \func{removeKnownPeer(pk)} & Runtime $\to$ Network & Withdraw recognition and recorded address \\
\midrule
\multicolumn{4}{l}{\textbf{Connection \& Reachability}} \\
\midrule
Connect & \func{onPeerConnected(pk, underlay, application?)} & Runtime $\leftarrow$ Network & Peer reachable on an underlay; pk, underlay, and the attested application, or null for an unattested identity \\
\rowcolor{rowgray}
Disconnect & \func{onPeerDisconnected(pk, underlay)} & Runtime $\leftarrow$ Network & Peer unreachable on an underlay; queue messages \\
Reachable & \func{isPeerReachable(pk)} & Runtime $\to$ Network & Check before send \\
\rowcolor{rowgray}
Transports & \func{peerUnderlays(pk)} & Runtime $\to$ Network & Underlays now reaching pk (proximity) \\
\midrule
\multicolumn{4}{l}{\textbf{Communication}} \\
\midrule
Send & \func{send(pubKey, payload)} & Runtime $\to$ Network & Implements madGLP Send transaction \\
\rowcolor{rowgray}
Receive & \func{onMessageReceived(sender, payload, messageId, underlay)} & Runtime $\leftarrow$ Network & Feeds madGLP Receive transaction \\
\midrule
\multicolumn{4}{l}{\textbf{Cryptography}} \\
\midrule
Sign & \func{sign(bytes)} & Runtime $\to$ Network & Ed25519 signature under the agent's key \\
\rowcolor{rowgray}
Verify & \func{verify(pk, bytes, signature)} & Runtime $\to$ Network & Check an Ed25519 signature by a public key \\
\midrule
\multicolumn{4}{l}{\textbf{Proximity underlays (PAN and LAN)}} \\
\midrule
\rowcolor{rowgray}
Scan & \func{\{pan,lan\}.\allowbreak{}getPeers()} & Runtime $\to$ Network & Find nearby agents on that underlay \\
Notify & \func{\{pan,lan\}.\allowbreak{}onPeerDiscovered(pk)} & Runtime $\leftarrow$ Network & Nearby Grassroots agent on that underlay \\
\rowcolor{rowgray}
Policy & \func{\{pan,lan\}.\allowbreak{}setTrustLevel(level)} & Runtime $\to$ Network & Set that underlay's cold-call trust level \\
\midrule
\multicolumn{4}{l}{\textbf{WAN}} \\
\midrule
\rowcolor{rowgray}
Connectivity & \func{wan.\allowbreak{}onConnectivityStatusChanged\allowbreak{}(old?, new?)} & Runtime $\leftarrow$ Network & Public IP change; (old?, new?) \\
Own address & \func{wan.\allowbreak{}getPublicAddress()} & Runtime $\leftarrow$ Network & Current public ip:port, share with peers \\
\rowcolor{rowgray}
Local network & \func{wan.\allowbreak{}networkIdentity()} & Runtime $\leftarrow$ Network & Fingerprint of the attached local network \\
Network change & \func{wan.\allowbreak{}onNetworkChanged(oldId?, newId?)} & Runtime $\leftarrow$ Network & Local network changed; (old?, new?) \\
\rowcolor{rowgray}
Feed address & \func{wan.\allowbreak{}putPeerAddress(pk, address)} & Runtime $\to$ Network & Hand the layer a GLP-learned peer address \\
Observed address & \func{wan.\allowbreak{}observedPeerAddress(pk)} & Runtime $\to$ Network & Public address at which a connected peer is observed \\
\rowcolor{rowgray}
Punch & \func{wan.\allowbreak{}punchUdp(address)} & Runtime $\to$ Network & UDP hole-punch toward an address \\
Declare server & \func{wan.\allowbreak{}putRendezvousServer(pk, address)} & Runtime $\to$ Network & A rendezvous server the layer connects to \\
\rowcolor{rowgray}
Remove server & \func{wan.\allowbreak{}removeRendezvousServer(pk)} & Runtime $\to$ Network & Withdraw it \\
\midrule
\multicolumn{4}{l}{\textbf{Places}} \\
\midrule
Declare place & \func{declarePlace(place, radius)} & Runtime $\to$ Network & Register a geofence about the device's location \\
\rowcolor{rowgray}
Remove place & \func{removePlace(place)} & Runtime $\to$ Network & Unregister it \\
Place events & \func{onPlaceEvent(place, event)} & Runtime $\leftarrow$ Network & Crossings, and when they cannot be observed \\
\bottomrule
\end{longtable}

\section*{Acknowledgments}
We thank Idit Keidar and Ohad Eitan for the productive discussions, and the helpful comments and reviews they provided to make this paper.

\bibliographystyle{plain}
\bibliography{bib}

@article{shapiro2026implementing,
  title={Implementing Grassroots Logic Programs with Multiagent Transition Systems and AI (Full Version)},
  author={Shapiro, Ehud},
  journal={arXiv:2602.06934.  Summary to appear in Proc. of LOPSTR+PPDP'26},
  year={2026},
  url = {https://arxiv.org/abs/2602.06934}
}

@article{shapiro2025glp,
  title={GLP: A Grassroots, Multiagent, Concurrent, Logic Programming Language for AI (full version)},
  author={Shapiro, Ehud},
  journal={arXiv preprint arXiv:2510.15747, Summary in Proc. of ICLP'26, EPTCS 450, pp. 119--133},
  year={2025},
  url = {https://arxiv.org/abs/2510.15747}
}

@inproceedings{shapiro2023grassrootsBA, 
author={Shapiro, Ehud},
 title={Grassroots Distributed Systems: Concept, Examples, Implementation and Applications (Brief Announcement)
},year = {2023},
publisher = {LIPICS},
booktitle = {37th International Symposium on Distributed Computing (DISC 2023). (Extended version: arXiv:2301.04391)},
address = {Italy},
notes={Extended version: arXiv preprint arXiv:2301.04391},
pages = {47:1--47:7},
  url = {https://arxiv.org/abs/2301.04391}
}

@inproceedings{shapiro2025atomic,
  title={Grassroots Platforms with Atomic Transactions: Social Graphs, Cryptocurrencies, and Democratic Federations},
  author={Shapiro, Ehud},
  booktitle={Proceedings of the 27th International Conference on Distributed Computing and Networking},
  pages={71--81},
  doi = {10.1145/3772290.3772309},
  note = {arXiv preprint arXiv:2502.11299},
  year={2026}
}

@misc{appleAppIntegrity,
  author       = {{Apple}},
  title        = {{Establishing your app's integrity}},
  howpublished = {Apple Developer Documentation, \url{https://developer.apple.com/documentation/devicecheck/establishing-your-app-s-integrity}},
  year         = {2026},
  note         = {Accessed 22 October 2026}
}

@article{eitan2026securing,
      title={Securing People and their Machines Against Major Faults (Full Version)}, 
      author={Ohad Eitan and Idit Keidar and Ehud Shapiro},
  journal={Proceedings of The 28th International Symposium on Stabilization, Safety, and Security of Distributed Systems (SSS 2026), to appear},
  note={arXiv:2607.02304},
  year={2026},
  url = {https://arxiv.org/abs/2607.02304}
}

@misc{rfc8445,
  author       = {Ari Keranen and Christer Holmberg and Jonathan Rosenberg},
  title        = {Interactive Connectivity Establishment ({ICE}): A Protocol for Network Address Translator ({NAT}) Traversal},
  howpublished = {RFC 8445},
  year         = {2018},
  url          = {https://www.rfc-editor.org/rfc/rfc8445},
}
\end{document}